\documentclass{ws-procs961x669}            % default, citations in superscript
\usepackage{graphicx}
\begin{document}
\title{Probing deviations to Kerr geometry with extreme mass-ratio inspirals}

\author{Shailesh Kumar}

\address{Indian Institute of Technology, Gandhinagar, Gujarat-382355, India\\
E-mail: shailesh.k@iitgn.ac.in}

\begin{abstract}
One of the primary research aims of the Laser Interferometer Space Antenna (LISA) mission is to comprehensively map the Kerr spacetime, a fundamental pursuit in the realm of general relativity. To achieve this goal, it is essential to develop precise tools capable of discerning any deviations from the Kerr geometry. Extreme mass-ratio inspirals (EMRIs) stand out as particularly promising sources for probing the spacetime metric, offering profound insights into gravitational phenomena. In this direction, we analyze a deformed Kerr geometry, being the central source of an EMRI system, with an inspiralling object that exhibits eccentric equatorial motion. We conduct a leading-order post-Newtonian analysis and examine the deviations in gravitational wave flux and phase with their leading-order contributions. Our findings evaluate the detectability of these deviations through gravitational wave dephasing and mismatch, highlighting the pivotal role of LISA observations in advancing our understanding of spacetime geometry.
\end{abstract}


\bodymatter

\section{Introduction}\label{aba:sec1}
Gravitational waves (GWs), envisioned by Einstein's general theory of relativity, stand as a groundbreaking avenue for observational astronomy and astrophysics, offering novel opportunities for investigating the universe's most intriguing and dynamic phenomena \cite{LIGOScientific:2016aoc, LIGOScientific:2016sjg, LIGOScientific:2017bnn}. It presents an unparalleled opportunity to thoroughly examine deviations from general relativity (GR), enabling the systematic exploration of alternative theories of gravity and the precise testing of their theoretical investigations. Driven by such a motivation, binary black hole systems, in particular, \textit{extreme mass-ratio inspirals}—a binary system with an inspiralling stellar-mass object (secondary, mass $\mu$) and a supermassive black hole (primary, mass $M$) whose mass-ratio ($q$) lies in the range of $q\equiv \mu/M=10^{-7}-10^{3}$—offers an invaluable chance to test GR and beyond in the extreme regime of gravity with low-frequency detectors. Such sources can serve as a potential probe to examine deviations from GR \cite{Babak:2017tow, Amaro-Seoane:2007osp, Rahman:2022fay, Zi:2023qfk, Maselli:2021men, Gair:2017ynp, LISA:2022kgy, Rahman:2023sof, Gair:2011ym, Zhang:2024csc, Moore:2016qxz, Ghosh:2024arw}. 

Besides the motivation coming from cosmological implications, GR has been tested in various aspects, but extreme conditions such as those near black holes, neutron stars, or during the early universe may not be fully probed. Modified gravity theories can predict deviations from GR in these extreme conditions, providing a way to test the robustness of GR. To investigate the observational consequences of such theories, EMRIs may serve as a useful tool to probe strong gravity regimes. Researchers follow various technical approaches to investigate EMRI sources; for example, one may use black hole perturbation or post-Newtonian frameworks. For weak-field regime analysis, the post-Newtonian (PN) approach is often sufficient \cite{Kumar:2024utz, AbhishekChowdhuri:2023gvu, Zi:2023qfk}, which is also of interest to this article. A feasible approach is to use a deformed Kerr metric, where deviations are described by parametric modifications \cite{Johannsen:2013szh, Yagi:2023eap, Kumar:2024utz, AbhishekChowdhuri:2023gvu}. Several frameworks have been developed to investigate potential observational signatures of black holes that deviate from the standard Kerr model\cite{Collins:2004ex, PhysRevD.81.024030}. We focus on the inspiral phase of a binary system and employ the leading-order PN framework to examine beyond GR signatures. 

In this direction, we consider a general axisymmetric, stationary, and asymptotically flat spacetime that brings deviations to Kerr geometry and presents the orbital dynamics in section (\ref{sec2}) \cite{Johannsen:2013szh, Yagi:2023eap}. The deviation parameters can further be mapped to the theories beyond GR, thus providing a useful and general setup to analyze distinct theories. In section (\ref{sec3}), we investigate such a spacetime with and without parametrized post-Newtonian (ppN) constraints, analyzing the emergence of deviation parameters in fluxes with their corresponding post-Newtonian (PN) orders. We further provide an order of magnitude analysis by computing dephasing and mismatch in sections (\ref{sec4}), implying promising detectability from the Laser Interferometer Space Antenna (LISA) observations.  

\section{Deformed Kerr geometry and orbital motion}\label{sec2}
We know that the Kerr black hole has received unprecedented success. Let us pose a question: how do we bring the deformations (deviations) in the Kerr metric in a more generic sense? To address such an aspect, we use the Johannsen framework \cite{Johannsen:2013szh, Yagi:2023eap}, where he came up with the idea of adding deformations in a perturbative manner in the contravariant components of the Kerr metric such that the geodesic equations are still separable in radial and angular parts, i.e., the Hamilton-Jacobi equation is separable. There are two cases we briefly discuss here. 
\vspace{-0.2cm}
\subsection{Johannsen (JHN) spacetime}\label{JHN}
In this section, we present the spacetime that Johannsen explored to bring deviations to Kerr geometry, which has the symmetries of Kerr black holes and three constants of motion: energy, angular momentum, and Carter constant. This is considered to be a generic stationary, axisymmetric and asymptotically flat metric, giving the notion of non-GR deviations. The related details of the metric can be found in  \cite{Johannsen:2013szh, Staelens:2023jgr}. The metric is written as:
\begin{align}\label{metric}
ds^{2} =& -\tilde{\Sigma}\frac{\Delta-a^{2}A_{2}^{2}\sin^{2}\theta}{N}dt^{2}+\frac{\tilde{\Sigma}}{\Delta A_{5}}dr^{2}+\tilde{\Sigma} d\theta^{2}-2a\tilde{\Sigma}\sin^{2}\theta \frac{(r^{2}+a^{2})A_{1}A_{2}-\Delta}{N}dt d\phi + \nonumber \\ 
& \tilde{\Sigma} \sin^{2}\theta\frac{(r^{2}+a^{2})^{2}A_{1}^{2}-a^{2}\Delta\sin^{2}\theta}{N}d\phi^{2}, 
\end{align}
where $A_{1}, A_{2}, A_{5}$ and $N$ are functions of radial coordinate $r$, given by
\begin{align}
A_{1}(r) =& 1+\alpha_{13}\Big(\frac{M}{r}\Big)^{3} \hspace{7mm} ; \hspace{7mm} N(r) = \Big((r^{2}+a^{2})A_{1}-a^{2}A_{2}\sin^{2}\theta\Big)^{2}\,,\nonumber\\
A_{2}(r) =& 1+\alpha_{22}\Big(\frac{M}{r}\Big)^{2} \hspace{7mm} ; \hspace{7mm} \Delta(r) = r^{2}-2M r+a^{2}\,, \nonumber\\
A_{5}(r) =& 1+\alpha_{52}\Big(\frac{M}{r}\Big)^{2} \hspace{7mm} ; \hspace{7mm} \tilde{\Sigma} = r^{2}+a^{2}\cos^{2}\theta+\epsilon_{3} \frac{M^{3}}{r}, \label{eq2}
\end{align}
where ($\alpha_{13}, \alpha_{22}, \alpha_{52}, \epsilon_{3}$) denote deformation parameters in Kerr spacetime. We will often denote the results in subsequent sections by a subscript \textit{JHN} (Johannsen). The interesting point about this metric is that the event horizon and the multipole moments coincide with Kerr's. It is important to note that the ppN (parametrized post-Newtonian) constraints are derived under the assumption that the central object is a star. Consequently, to apply these constraints, we must further assume that the asymptotic behaviours of the metric components for both star and black hole geometries are identical \cite{Cardoso:2014rha, Carson:2020dez}. Further, we consider the leading-order post-Newtonian (PN) analysis to examine the behaviour of deviations to GR. We take small values of deviation parameters as we perform the analysis in a perturbative manner. This implies the leading-order contribution of deviations in the GW dephasing. 

%\subsection{Orbital motion}
We consider that a test object exhibits eccentric equatorial motion in the Johannsen background. As mentioned earlier, the spacetime (\ref{metric}) has three constants of motion: energy ($E$), angular momentum ($J_{z}$) and Carter constant ($\mathcal{Q}$). Since the motion is equatorial, this implies $\mathcal{Q}=0$. Also, for computational convenience, we perform our analysis in dimensionless units: $\hat{r}=r/M, \hat{a}=a/M, \hat{J}_{z}=J_{z}/(\mu M)$. In the end, one can always dimensionalize the observable in physical units. With this, we express geodesic velocities of the inspiralling object using the Hamilton-Jacobi method; the action and corresponding equation of motion are given by,
\begin{align}\label{ac1}
S = -\frac{1}{2}\mu^{2}\tau-Et+J_{z}\phi+R(r)+\Theta(\theta) \hspace{3mm} ; \hspace{3mm} -\frac{\partial S}{\partial\tau} = \frac{1}{2}g^{\mu\nu}\frac{\partial S}{\partial x^{\mu}}\frac{\partial S}{\partial x^{\nu}}\,.
\end{align}
Following \cite{AbhishekChowdhuri:2023gvu}, we replace $E=\mu+\mathcal{E}$ for separating out the rest-mass energy; consequently, we reach the following expressions with leading-order corrections in deviations and black hole spin,
% \begin{equation}
% \begin{aligned}\label{gdscs1}
% \frac{dt}{d\tau}  =& \frac{1}{A_{5} \Sigma}(A_{1}^2 E+a J_{z} (A_{5}-A_{0}))\,, \\
% \frac{d\phi}{d\tau} =& \frac{1}{A_{5} \Sigma}(A_{5} J_{z} \csc ^2\theta-a \left(A_{5}-A_{0}\right)E)\,, \\
% \Big(\frac{dr}{d\tau}\Big)^{2} =& \frac{A_{5}}{\Sigma^{2}}\Big[-C-\mu^{2}(f+r^{2})-\frac{1}{A_{5}}(-a^{2}J_{z}^{2}A_{2}^{2}+2aEJ_{z}A_{0}-E^{2}A_{1}^{2}) \Big]\,, \\
% \Big(\frac{d\theta}{d\tau}\Big)^{2} =& \frac{1}{\Sigma^{2}}\Big[C-(a^{2}E^{2}\sin^{2}\theta+\mu^{2}a^{2}\cos^{2}\theta-2aEJ_{z}+J_{z}^{2}\csc^{2}\theta) \Big]\,,
% \end{aligned}
% \end{equation}
\begin{equation}
\begin{aligned}\label{gdscs3n}
\Big(\frac{dr}{d\tau}\Big)^{2} = & 2 \mathcal{E}+\frac{2}{r}-\frac{J_{z}^{2}}{r^{2}}+-\frac{4aJ_{z}}{r^{3}}+\frac{2\alpha_{13}}{r^{3}}-\frac{\epsilon_{3}}{r^{3}}\Big(1-\frac{2J_{z}^{2}}{r^{2}}\Big)+\frac{\alpha_{52}}{r^{3}}\Big(2-\frac{J_{z}^{2}}{r}\Big) \\
\frac{d\phi}{d\tau} =& \frac{J_{z}}{r^2}+\frac{2a}{r^{3}}-\frac{\epsilon_{3}J_{z}}{r^{5}}.
\end{aligned}
\end{equation}
One can further use the radial part in Eq. (\ref{gdscs3n}) to derive an expression for the last stable orbit (LSO), also termed as \textit{separatrix} ($p_{sp}$), which will be governed by the following equation:
\begin{align}\label{vef1}
V_{eff}(r) \equiv -\frac{1}{2}(g^{tt}E^{2}-2g^{t\phi}EJ_{z}+g^{\phi\phi}J_{z}^{2}+1)=\Big(\frac{dr}{d\tau}\Big)^{2}\,.
\end{align}
As the object exhibits the eccentric motion, there are two turning points: \textit{periastron} ($r_{p}$) and \textit{apastron} ($r_{a}$), within which the object will show the bounded orbits with the semi-latus rectum ($p$) and eccentricity ($e$): $r_{p}=p/(1+e)$ and $r_{a}=p/(1-e)$. Solutions for such orbits can be obtained in the range $r_{p}<r<r_{a}$ if $V_{eff}(r)<0\,.$ As a result, the LSO takes the following form \cite{AbhishekChowdhuri:2023gvu}
\begin{equation}
\begin{aligned}\label{sp1}
p_{sp}=& 2(e+3)-4a\sqrt{2}\sqrt{\frac{1+e}{3+e}}+4\alpha_{13}\frac{(1+e)}{(3+e)^{2}}-\epsilon_{3}\frac{(e-3)^{2}(e+1)^{2}}{4(e+e)^{3}},
\end{aligned}
\end{equation}
where if we switch off the deviations, the Schwarzschild results are recovered. It is to note that we are only focusing on the leading-order contributions of deviation parameters with ppN constraints employed. We use the LSO to truncate the trajectory of the inspiralling object. We will also find at the later stage that the leading-order PN analysis with the contributions of leading-order deviations in the observable emerges at the 2PN.

\subsection{A more generic picture of spacetime by YLLC}\label{sec2}
A recent study describes a more general picture with deformations where the Johannsen spacetime becomes a subclass. Such a spacetime initially was given by Carson-Yagi (CY) \cite{Carson:2020dez}, which was later on improved by CY and their colleagues \textit{Kent Yagi, S. Lomuscio, T. Lowrey, Z. Carson}\cite{Yagi:2023eap} (YLLC), in order to remove the unphysical divergences or pathological behaviour in the Johannsen and CY metric. The metric is given by\cite{Kumar:2024utz}
\begin{equation}
\begin{aligned}\label{mets}
ds^{2} =& -\frac{\tilde{\Sigma}A_{5}}{\rho^{4}}(A_{5}-a^{2}A_{2}^{2}\sin^{2}\theta)dt^{2}+\frac{\tilde{\Sigma}}{A_{5}}dr^{2}+\tilde{\Sigma}d\theta^{2}+\frac{2aA_{5}\tilde{\Sigma}}{\rho^{4}}(A_{5}-A_{0})\sin^{2}\theta dtd\phi + \\  
& \frac{\tilde{\Sigma}A_{5}\sin^{2}\theta}{\rho^{4}}(A_{1}^{2}-a^{2}A_{5}\sin^{2}\theta)d\phi^{2}, 
\end{aligned}
\end{equation}
where $\rho^{4} = a^{4}A_{2}^{2}A_{5}\sin^{4}\theta+a^{2}\sin^{2}\theta (A_{0}^{2}-2A_{0}A_{5}-A_{1}^{2}A_{2}^{2})+A_{1}^{2}A_{5}$, and
$\Tilde{\Sigma} \equiv \Sigma + f(r) + g(\theta) \hspace{1mm} ; \hspace{1mm} \Sigma \equiv r^{2}+a^{2}\cos^{2}\theta \hspace{1mm} ; \hspace{1mm} \Delta \equiv r^{2}+a^{2}-2Mr$. Without loss of generality, one can consider $g(\theta)=0$. The deviation functions are given by\cite{Kumar:2024utz}
\begin{equation}\label{dfgn}
\begin{aligned}
A_{0} =& r^{2}\Big(1+\frac{a^{2}}{r^{2}}+\alpha_{01} \Big(\frac{M}{r}\Big)) \hspace{3mm} ; \hspace{3mm}
A_{1} = r^{2}\Big(1+\frac{a^{2}}{r^{2}}+\alpha_{11} \Big(\frac{M}{r}\Big)+\alpha_{12}\Big(\frac{M}{r}\Big)^{2}\Big)\,, \\
A_{2} =& 1+\alpha_{21}\Big(\frac{M}{r}\Big) \hspace{1mm} ; \hspace{1mm}
A_{5} = r^{2}\Big(1-\frac{2M}{r}+\frac{a^{2}}{r^{2}}+\alpha_{51}\Big(\frac{M}{r}\Big)\Big)\hspace{1mm} ; \hspace{1mm}
f(r) = r^{2}\Big(\epsilon_{1}\frac{M}{r}\Big)\,,
\end{aligned}
\end{equation}
where ($\alpha_{01}, \alpha_{11},  \alpha_{12}, \alpha_{21}, \alpha_{51}, \epsilon_{1}$) are deformations to Kerr geometry. If we take $A_{0}\longrightarrow A_{1}A_{2}$, one can easily recover the Johannsen spacetime. Again, this is a more generic picture of deformed Kerr geometry, which is stationary, axisymmetric, asymptotically flat and contains three constants of motion similar to the case of Johannsen. 

Note that we have not taken the ppN constraints into consideration for this particular case. Consequently, upon performing the leading-order PN analysis, we will see that the deviation parameters contribute to the 0PN itself. However, if we put ppN constraints, we can easily see the contributing deviations are ($\alpha_{13}, \alpha_{52}, \epsilon_{3}$). One can always compute the higher order corrections up to 2PN and beyond. One has to derive all relevant expressions that involve deformations contributing up to 2PN, including the GR results. Such an analysis requires further advancements in the methodology, which would be communicated in a future study. For comprehensive details, readers are suggested to look at \cite{Kumar:2024utz, AbhishekChowdhuri:2023gvu, Canizares:2012is}. Therefore, we will address observational characteristics with leading-order PN analysis only in both of the scenarios separately, i.e., Johannsen and VLLC pictures.

Now with the use of metric (\ref{mets}), following the analysis of section (\ref{JHN}), we obtain,
\begin{equation}
\begin{aligned}\label{gdscs3n1}
\Big(\frac{dr}{d\tau}\Big)^{2} = & 2 \mathcal{E}+\frac{1}{r}(2+2 \alpha_{11}-\alpha_{51}-\epsilon_{1})-\frac{1}{r^{2}}J_{z}^{2} \hspace{3mm} ; \hspace{3mm}
\frac{d\phi}{d\tau} = \frac{J_{z}}{r^2}+\frac{2a}{r^3}-J_{z}\frac{\epsilon_{1}}{r^3}\,.
\end{aligned}
\end{equation}
Further, with the use of the radial expression in Eq. (\ref{gdscs3n1}), we obtain the LSO of the following form
\begin{equation}
\begin{aligned}\label{sp1}
p_{sp}=& 2(e+3)+\frac{8(1+e)}{(3+e)^{2}}\alpha_{12}-(e+3)\alpha_{51}+ \frac{(3+2e-e^{2})}{(3+e)}\epsilon_{1}\,.
\end{aligned}
\end{equation}
Note that the dominant contribution of deviations emerges at the 0PN. Hence, higher-order PN effects will no longer contribute. Next, we provide a brief setup with radiation reaction effects for computing the average fluxes generated during the motion of the inspiralling object. 

\section{Radiation reaction: fluxes, orbital evolution}\label{sec3}
In this section, we proceed to compute the observable quantities in both the cases for Johannsen spacetime (with ppN constraints) and YLLC metric (without ppN constraints). We consider radiation reaction on the inspiralling object, as a result, we estimate GW fluxes and the orbital evolution of the dynamics with the effects of deviations. Since we are considering an EMRI system, the inspiralling object has very little mass with respect to the central supermassive black hole. This implies a perturbative system where the secondary object of the EMRI system perturbs the background of the primary. This gives rise to the notion of radiation reaction, implying the constants of motion ($\mathcal{E}, J_{z}$) are no longer constants. Instead, they evolve in time. With this, we compute the leading-order effects of deviations on average loss of energy and angular momentum that further provide the evolution of the secondary object \cite{Flanagan:2007tv, PhysRevD.52.R3159, Ryan:1995xi}. 

We further introduce a useful radial parametrization that helps remove diverging behaviour in differential equations at turning points. The parameterization is given by
\begin{align}
    r = \frac{p}{1+e \cos\chi}
\end{align}
We will use this parameterization to calculate the dynamics in terms of semi-latus rectum ($p$) and eccentricity ($e$). We start the computation of fluxes by representing the quantities in terms of Cartesian coordinates: ($x_{1},x_{2},x_{3})=(r\sin\theta \cos\phi, r\sin\theta \sin\phi, r\cos\theta$), which helps computationally in writing the geodesic velocities. Since we are focusing on an equatorial plane ($\theta=\pi/2$), the instantaneous fluxes can be written as \cite{AbhishekChowdhuri:2023gvu, Kumar:2024utz}
\begin{equation}
\begin{aligned}\label{inst fluxes}
\dot{\mathcal{E}}\vert_{JHN} =& x_{i}\Ddot{x}_{i} \hspace{5mm} ; \hspace{5mm}
\dot{J}_{z}\vert_{JHN} = \epsilon_{3jk}x_{j}\Ddot{x}_{k}\Big(1+\frac{\epsilon_{3}}{(x_{i}x_{i})^{3/2}}\Big) \\
\dot{\mathcal{E}}\vert_{YLLC} =& x_{i}\Ddot{x}_{i} \hspace{5mm} ; \hspace{5mm}
\dot{J}_{z}\vert_{YLLC} = \epsilon_{3jk}x_{j}\Ddot{x}_{k}\Big(1+\frac{\epsilon_{1}}{\sqrt{x_{i}x_{i}}}\Big)
\end{aligned}
\end{equation}
We notice that the Eq. (\ref{inst fluxes}) involves a double derivative of $x_{i}$, representing the contribution of acceleration which we often denote as $a_{j}$. This is also called as \textit{radiation reaction acceleration}. The general expression of $a_{j}$ is written in terms of mass ($I_{jk}$) and current ($J_{jk}$) quadrupole moments, given as\cite{Flanagan:2007tv, PhysRevD.52.R3159}
\begin{align}\label{accelration}
a_j=-\dfrac{2}{5}I^{(5)}_{jk}x_{k}+\dfrac{16}{45}\epsilon_{jpq}J^{(6)}_{pk}x_{q}x_{k}+\dfrac{32}{45}\epsilon_{jpq}J^{(5)}_{pk}x_{k} \dot{x}_{q}+\dfrac{32}{45}\epsilon_{pq[j}J^{(5)}_{k]p}x_{q} \dot{x}_{k}+\dfrac{8J}{15}J^{(5)}_{3i},
\end{align}
where,
\begin{equation}\label{moments}
I_{jk}=\Big[x_j x_k\Big]^{\text{STF}} \hspace{3mm} ; \hspace{3mm}
J_{jk}=\Big[x_{j}\epsilon_{kpq}x_{p}\dot{x}_{q}-\dfrac{3}{2}x_jJ\delta_{k3}\Big]^{\text{STF}}.
\end{equation}
where ($I_{jk}, J_{jk}$) are the symmetric trace-free (STF) contributions. $B_{[ij]}$ is an anti-symmetric quantity: $B_{[ij]}=\frac{1}{2}(B_{ij}-B_{ji})$. The superscripts denote the derivative order, and $J$, in the last term, refers to the black hole spin $a$. Finally, we use the set of Eqs. (\ref{moments}, \ref{accelration}) for computing the rate change of constants of motion represented in Eq. (\ref{inst fluxes}). 

Our goal is to estimate the average rate change of constants of motion ($<\dot{\mathcal{E}}>, <\dot{J_{z}}>$). We consider adiabatic approximation \cite{Hinderer:2008dm, PhysRevD.103.104014, PhysRevLett.128.231101, Glampedakis:2002ya}, where the dynamics of the particle are approximated by geodesics. Alternatively, we assume that the temporal scales are significantly shorter than the radiation reaction time scale. This implies slow changes in instantaneous fluxes. Since the object exhibits eccentric motion, one should average out the instantaneous quantities over a period, given as
\begin{align}\label{sdg}
<\Dot{\mathcal{E}}> = \frac{1}{T_{r}}\int_{0}^{2\pi}\Dot{\mathcal{E}}\frac{dt}{d\chi}d\chi \hspace{3mm} ; \hspace{3mm} <\Dot{J_{z}}> = \frac{1}{T_{r}}\int_{0}^{2\pi}\Dot{J_{z}}\frac{dt}{d\chi}d\chi,
\end{align}
where $\chi\in (0,2\pi)$. The radial time period for the spacetimes under consideration are
\begin{equation}
\begin{aligned}\label{erg}
T_{r}\vert_{JHN} =& \frac{\pi}{\sqrt{1-e^{2}}}\Big[\frac{2p^{3/2}}{1-e^{2}}-6a+\frac{1}{\sqrt{p}}(3\alpha_{13}-\alpha_{52}-3\epsilon_{3}/2) \Big] \\
T_{r}\vert_{VLLC} =& \frac{2\pi p^{3/2}}{(1-e^{2})^{3/2}}\Big(1-\frac{\alpha_{11}}{2}+\frac{\alpha_{51}}{4}+\frac{\epsilon_{1}}{4}\Big)\,.
\end{aligned}
\end{equation}
Hence, using Eqs. (\ref{gdscs3n}, \ref{gdscs3n1}, \ref{sdg}, \ref{erg}), we obtain the following average loss of energy and angular momentum fluxes for the case of Johannsen spacetime, mentioned in section (\ref{JHN}) \cite{AbhishekChowdhuri:2023gvu}
\begin{equation}
\begin{aligned}\label{avgflx1}
\Big\langle\frac{d\mathcal{E}}{dt}\Big\rangle\Big\vert_{JHN} =& -\frac{(1-e^2)^{3/2}}{15 p^{5}}\Big[(37 e^4+292 e^2+96)-\Big\{\frac{a}{p^{3/2}} (\frac{491 e^6}{2}+2847 e^4 \\
& +3292 e^2+584) -\frac{\alpha_{13}}{p^2}(251 e^6+4035 e^4+5532 e^2+864)\\
& +\frac{\alpha_{52}}{p^2} \Big(\frac{2775 e^6}{4}+2275 e^4-130 e^2 -98e^{2} \sqrt{1-e^2} -48 \sqrt{1-e^2} \\
& +\frac{37e^{6} \sqrt{1-e^2}}{2}+\frac{255e^{4} \sqrt{1-e^2}}{2}-240\Big) \\
& +\frac{\epsilon_{3}}{p^{2}} \Big(\frac{251 e^6}{2} +\frac{4035 e^4}{2}+2766 e^2 + 432\Big)\Big\}\Big],
\end{aligned}
\end{equation}
\begin{equation}
\begin{aligned}\label{avgflx2}
\Big\langle\frac{dJ_{z}}{dt}\Big\rangle\Big\vert_{JHN} =& -\frac{(1-e^{2})^{3/2}}{5p^{3}}\Big[\frac{4}{p^{1/2}}\left(7 e^2+8\right)-\frac{a}{3 p^{2}} \left(549 e^4+1428 e^2+488\right) \\
& +\frac{2\alpha_{13}}{p^{5/2}} (84 e^4+361 e^2+120)-\frac{\alpha_{52}}{2(1-e^{2})^{1/2}p^{5/2}} \Big\{(-28 e^6+24 e^4 \\
& +36 e^2-32) +\sqrt{1-e^2}(12 e^6+421 e^4+192 e^2-128)\Big\} \\
& -\frac{\epsilon_{3}}{p^{5/2}} \Big(84 e^4 + 361e^2 + 120\Big)\Big].
\end{aligned}
\end{equation}
Similarly, for the case of a more generic picture of deformed Kerr geometry described in section (\ref{sec2})

\begin{equation}
\begin{aligned}\label{avflxn}
<\Dot{\mathcal{E}}>\Big\vert_{VLLC} =& -\frac{(1-e^{2})^{3/2}}{30p^{5}} (96+292e^{2}+37e^{4})(2+6\alpha_{11}-3\alpha_{51}-3\epsilon_{1}) \\
<\Dot{J}_{z}>\Big\vert_{VLLC} =& -\frac{(1-e^{2})^{3/2}}{5p^{7/2}}(8+7e^2)(4+10\alpha_{11}-5\alpha_{51}-5\epsilon_{1}).
\end{aligned}
\end{equation}
It is to be noted that if we restore the powers of the speed of light ($c$) in Eqs. (\ref{avgflx2}, \ref{avflxn}), and pull out an overall factor of $c^{-5}$; we can easily indicate the PN order as $c^{-2n}$, implying the order at which the deviation parameters contribute to the fluxes. As a result, we find that $<fluxes>\vert_{JHN} \sim c^{-5}(1+c^{-3}a+c^{-4} deviations)$ and $<fluxes>\vert_{VLLC} \sim c^{-5}(1 +deviations)$. This indicates that the leading-order deviations emerge at 2PN for the Johannsen spacetime when employing ppN constraints, whereas if we relax the ppN constraints and use the VLLC spacetime, we obtain the deviation at the 0PN itself \cite{AbhishekChowdhuri:2023gvu, Kumar:2024utz}. However, the contribution of black hole spin appears at 1.5PN \cite{Flanagan:2007tv}.

Since we are considering the adiabatic assumption, we can write down the flux-balance law \cite{AbhishekChowdhuri:2023gvu, Kumar:2024utz}, 
\begin{align}\label{blnceqn}
\Big\langle\frac{d\mathcal{E}}{dt}\Big\rangle_{GW} = -\Big\langle\frac{d\mathcal{E}}{dt}\Big\rangle \hspace{5mm} ; \hspace{5mm} \Big\langle\frac{dJ_{z}}{dt}\Big\rangle_{GW} = -\Big\langle\frac{dJ_{z}}{dt}\Big\rangle\,.
\end{align}
This implicates the generated GW fluxes. To calculate the GW flux, it’s essential to incorporate the metric perturbation into the deformed Kerr background. This can be expressed as $\sim g_{\mu \nu,\text{dKerr}}+h_{\mu \nu}$, where $g_{\mu \nu,\text{dKerr}}$
represents the background metric, and $h_{\mu\nu}$ denotes the perturbation due to GWs.  In this study, we focus solely on the corrections to the rates of change in orbital energy and angular momentum resulting from the deformation parameters. It is not obvious from the outset that the contributions to the GW flux from these deformation parameters will be subleading compared to the corrections to orbital energy and angular momentum. Investigating these contributions to the GW flux is not within the scope of this paper, and we plan to explore it in future work.

Next, we determine the orbital evolution of the inspiralling object using,
\begin{equation}
\begin{aligned}
\Big\langle\frac{dp}{dt}\Big\rangle = \Big(\frac{\dot{\mathcal{E}}\partial_{e}J_{z} -\dot{J}_{z}\partial_{e}\mathcal{E}}{\partial_{p}\mathcal{E}\partial_{e}J_{z}-\partial_{e}\mathcal{E}\partial_{p}J_{z}}\Big) \hspace{3mm} ; \hspace{3mm} \Big\langle\frac{de}{dt}\Big\rangle = \Big(\frac{\dot{J}_{z}\partial_{p}\mathcal{E}-\dot{\mathcal{E}}\partial_{p}J_{z}}{\partial_{p}\mathcal{E}\partial_{e}J_{z}-\partial_{e}\mathcal{E}\partial_{p}J_{z}}\Big)\,.
\end{aligned}
\end{equation}
and as a result, we obtain

\begin{equation}
\begin{aligned}\label{dpdtn}
\Big\langle\frac{dp}{dt}\Big\rangle\Big\vert_{JHN} =& -\frac{2(1-e^{2})^{3/2}}{5p^{3}}\Big[4(7 e^2+8)-\frac{a}{3 p^{3/2}}(475 e^4+1516 e^2+1064) \\
& +\frac{\alpha_{13}}{3 p^2}\left(1152+2384 e^2+509 e^4\right)\Big] +\frac{\alpha_{52} (1-e^2)^{3/2}}{15 p^5} (36 e^6+1273 e^4 \\
& -12 (7 e^2+8) (1-e^2)^{3/2}+4 e^2-672) \\
& +\frac{\epsilon_{3}(1-e^2)^{3/2}}{15 p^5} \Big[1152+2384 e^2+509e^{4}\Big] \\
\Big\langle\frac{de}{dt}\Big\rangle\Big\vert_{JHN} =& -\frac{e \left(1-e^2\right)^{3/2}}{15p^{4}}\Big[121 e^2+304-\frac{a}{2 p^{3/2}} (1313 e^4+5592 e^2+7032) \\
& +\frac{2 \alpha_{13}}{p^2} \left(385 e^4+2664 e^2+2292\right)\Big] +\frac{\alpha_{52}\left(1-e^2\right)^{3/2}}{60 e p^6} (72 e^8+5249 e^6 \\
& +6562 e^4 -2 e^2 (121 e^2+304) (1-e^2)^{3/2}-1872 e^2+384) \\
& +\frac{\epsilon_{3} e \left(1-e^2\right)^{3/2}}{60 p^6} \Big[9168+10656 e^2+1540 e^4\Big] \\
\Big\langle\frac{dp}{dt}\Big\rangle\Big\vert_{VLLC} =& -\frac{8(1-e^2)^{3/2}}{5p^{3}}(8+7e^{2})(1+2\alpha_{11}-\alpha_{51}-\epsilon_{1}) \\
\Big\langle\frac{de}{dt}\Big\rangle\Big\vert_{VLLC} =& -e\frac{(1-e^2)^{3/2}}{15p^4}(304+121e^{2}) (1+2\alpha_{11}-\alpha_{51}-\epsilon_{1})\,.
\end{aligned}
\end{equation}
It is obvious from the expressions in Eq. (\ref{dpdtn}) that if we integrate the above equation the evolution of orbital parameters ($p, e$) decreases over time. The main cause of this behaviour is that (\ref{dpdtn}) has negative leading terms on the right-hand sides, which guarantees a gradual decrease in both ($p, e$) over time. As long as the terms proportional to the deformation parameters continue to make subleading contributions, this decrease will continue. However, we avoided mentioning the numerical plots of the orbital evolution here as our main goal is to investigate the detectability of the deviations. Next, we examine such a study and perform an order of magnitude analysis to hunt for the non-GR deviations. 

\section{Detectability: GW dephasing and mismatch}\label{sec4}

As we are considering the inspiralling object exhibiting eccentric dynamics, it possesses two frequencies: angular frequency ($\Omega_{\phi}$) and radial frequency ($\Omega_{r}$). In principle, both frequencies execute their roles; however, we assume that the motion is governed by only azimuthal frequency ($\Omega_{\phi}$). This helps estimate the GW dephasing and corresponding order of magnitude for detecting non-GR deviations with LISA observations. The frequency is written as
\begin{align}
\frac{d\varphi_{i}}{dt} = \langle \Omega_{i}(p(t),e(t))\rangle = \frac{1}{T_{r}}\int_{0}^{2\pi}d\chi\frac{dt}{d\chi} \Omega_{i}(p(t),e(t),\chi) \hspace{5mm} ; \hspace{5mm} i = (\phi, r),
\end{align}
where $\langle\Omega_{i}\rangle$ denotes the averaged orbital frequency. The expression completely depends on ($p(t), e(t)$) that comes from Eq. (\ref{dpdtn}), apart from deviation parameters. We assume that the azimuthal frequency primarily regulates the analysis: $\varphi_{\phi}(t)\sim\phi (t)$ with $\varphi_{i}(0)=0$. Using Eq. (\ref{gdscs3n}), we obtain
\begin{equation}
\begin{aligned}\label{orbphase}
\frac{d\phi}{dt}\Big\vert_{VLLC} =& \frac{\left(1-e^2\right)^{3/2}}{p^{3/2}}-\frac{a \left(1-e^2\right)^{3/2}}{p^3}\left(3 e^2+1\right)+\frac{\alpha_{52} \left(1-e^2\right)}{4 p^{7/2}} (2 e^4- \\
& \left(3 \sqrt{1-e^2}+4\right) e^2+2) +\frac{3(1-e^{2})^{3/2}(1+e^{2})}{2p^{7/2}}\Big(\alpha_{13}-\frac{\epsilon_{3}}{2}\Big) \\
\frac{d\phi}{dt}\Big\vert_{JHN} =& \Big(\frac{1-e^2}{p}\Big)^{3/2}\Big(1+\frac{\alpha_{11}}{2}-\frac{\alpha_{51}}{4}-\frac{\epsilon_{1}}{4}\Big)\,.
\end{aligned}
\end{equation}
Again, if one restores the powers of the speed of light, we can see $d\phi/dt\vert_{JHN} \sim (1-c^{-3}a+c^{-4} deviations)$ and $d\phi/dt\vert_{VLLC} \sim c^{0}(1+deviations)$. This implies the contribution of deviations in GW phase at 2PN order in the case of Johannsen spacetime when employing ppN constraints, whereas the deviations emerge at the 0PN in the case when ppN constraints are relaxed in VLLC spacetime mentioned in section (\ref{sec2}).

\subsection{Dephasing}
Next, we estimate the effects of deviations on the GW phase and compute the GW dephasing. We integrate Eq. (\ref{orbphase}), with the use of Eq. (\ref{dpdtn}), to obtain the phase that includes the contribution of deviations and further subtract the GR part of the phase, which enables us to obtain the GW dephasing. This examination provides us with an order of magnitude for detecting non-GR deviations from LISA observations. As we are computing the quadrupolar results, we take the GW dephasing to be twice the orbital phase, i.e., $\Phi_{\textup{GW}} = 2\phi(t)$. The dephasing between a deformed Kerr geometry and Kerr black hole, up to a given time $t_{\textup{obs}}$, is written as
\begin{align}
\Delta\Phi(t_{\textup{obs}}) = \vert \Phi^{\textup{D-Kerr}}_{\textup{GW}}(t_{\textup{obs}}) - \Phi^{\textup{Kerr}}_{\textup{GW}}(t_{\textup{obs}}) \vert,
\end{align}
where $\Phi^{\textup{D-Kerr}}_{\textup{GW}}$ depicts the phase contributing from the Johannsen spacetime and $\Phi^{\textup{Kerr}}_{\textup{GW}}$ denotes the phase from the Kerr black hole. The difference between the two gives rise to the GW dephasing. We examine our study for the observation period of one year, $t_{\textup{obs}}=1$ year. Focusing on EMRIs, we take the primary black hole with the mass $M=10^{6}M_{\odot}$ and the secondary with the mass $\mu=10M_{\odot}$, sets the mass-ratio $q=10^{-5}$. We start the inspiral at $p_{\textup{in}}=14$ with different initial eccentricities ($e_{\textup{in}}$).
%%%%%%%%%%%%%%%%%%%%%%%%%%%%%%%%%%%%%%%%%%%%%%%%%%%%%%
\begin{figure}[htb!]
	%%%%%%%%%%%%%%%%%%%%%%%%
	\centering
	\minipage{0.33\textwidth}
	\includegraphics[width=\linewidth]{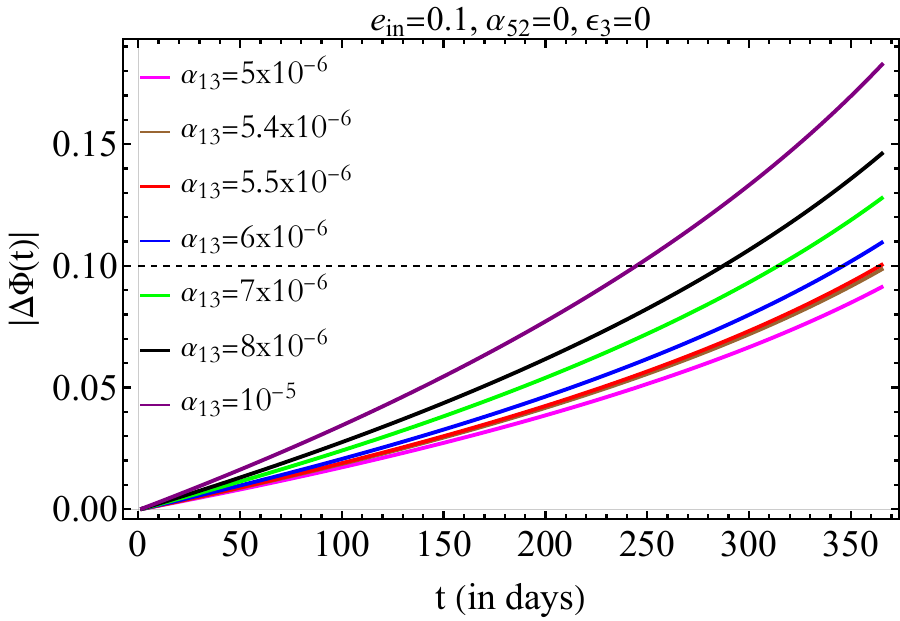}
% \caption{Wormholes for $\Lambda=0$}
	\endminipage\hfill
	%%%%%%%%%%%%%%%%%%%%%%%%
	\minipage{0.33\textwidth}
	\includegraphics[width=\linewidth]{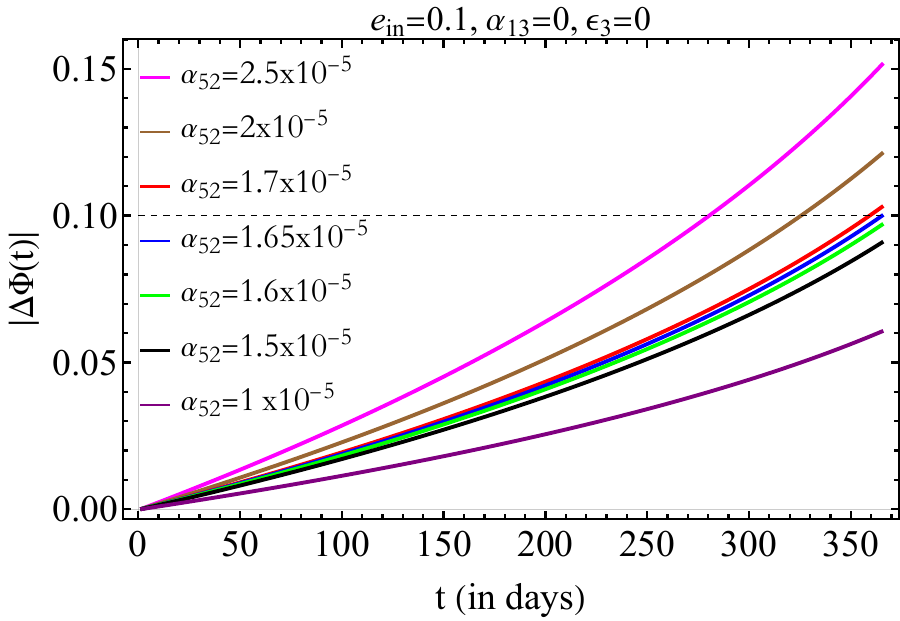}%{Dephasing_a0.05_alpha13_alpha52_epsilon3_eccen.pdf}
	\endminipage\hfill
  \minipage{0.33\textwidth}
         \includegraphics[width=\linewidth]{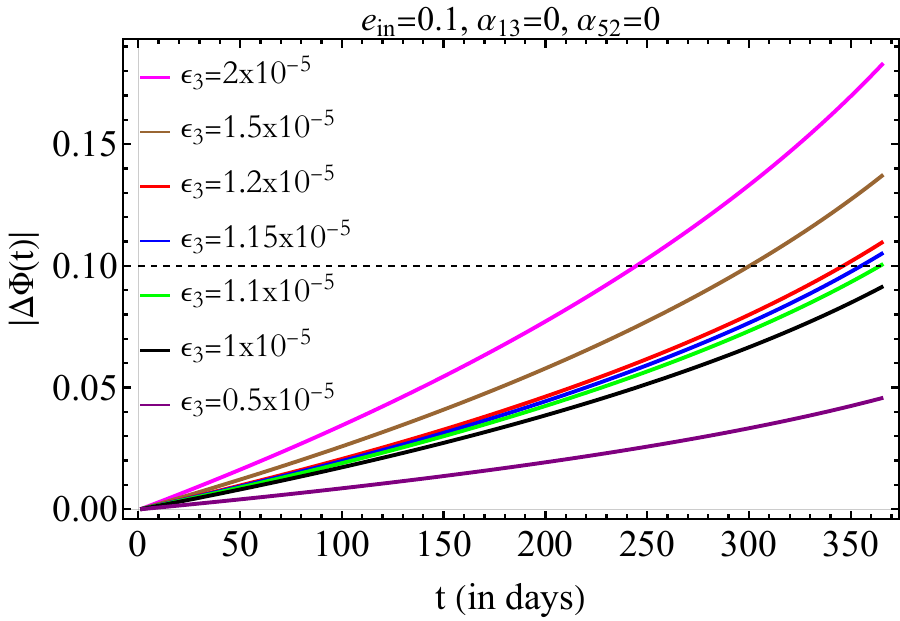}
	 \endminipage
	\caption{Dephasing for Johannsen spacetime (JHN): It provides an order of magnitude estimate of the upper bound on the deviation parameters when $|\Delta\Phi| <0.1$. 
 }\label{constrain_dephase1}
\end{figure}
%%%%%%%%%%%%%%%%%%%%%%%%%%%%%%%%%%%%%%%%%%%%%%%%%%%%%%%
\begin{figure}[htb!]
	%%%%%%%%%%%%%%%%%%%%%%%%
	\centering
	\minipage{0.33\textwidth}
	\includegraphics[width=\linewidth]{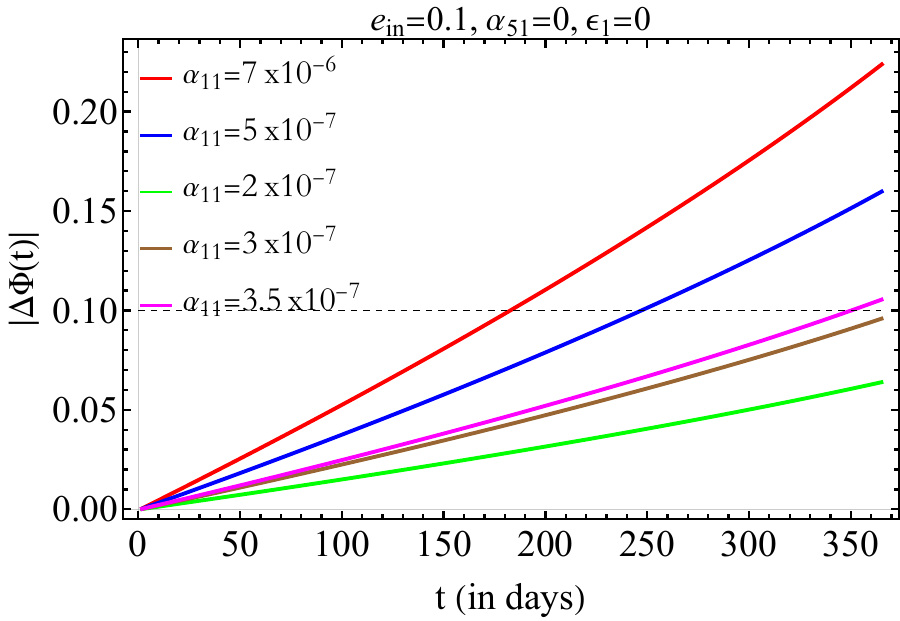}
% \caption{Wormholes for $\Lambda=0$}
	\endminipage\hfill
	%%%%%%%%%%%%%%%%%%%%%%%%
	\minipage{0.33\textwidth}
	\includegraphics[width=\linewidth]{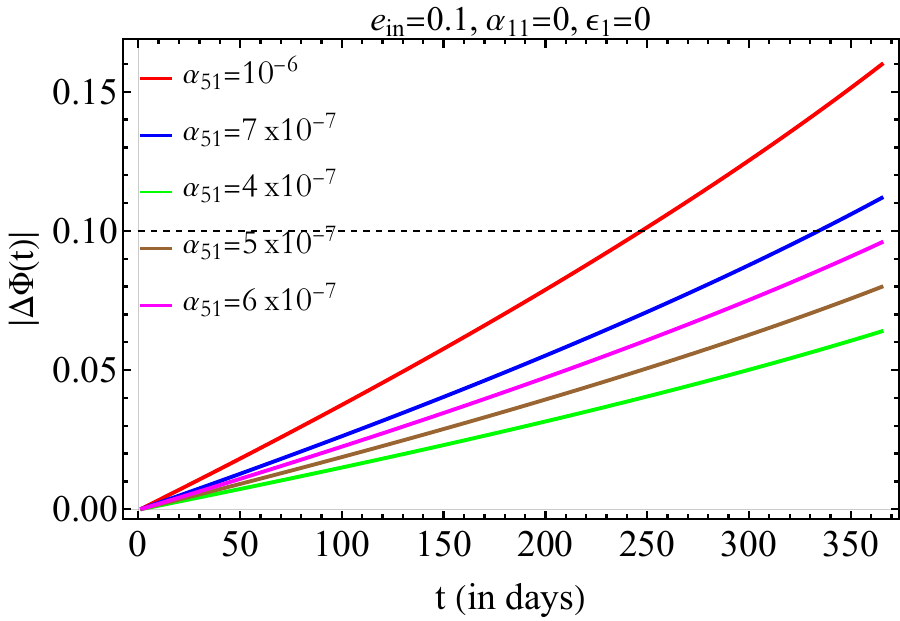}%{Dephasing_a0.05_alpha13_alpha52_epsilon3_eccen.pdf}
	\endminipage\hfill
  \minipage{0.33\textwidth}
         \includegraphics[width=\linewidth]{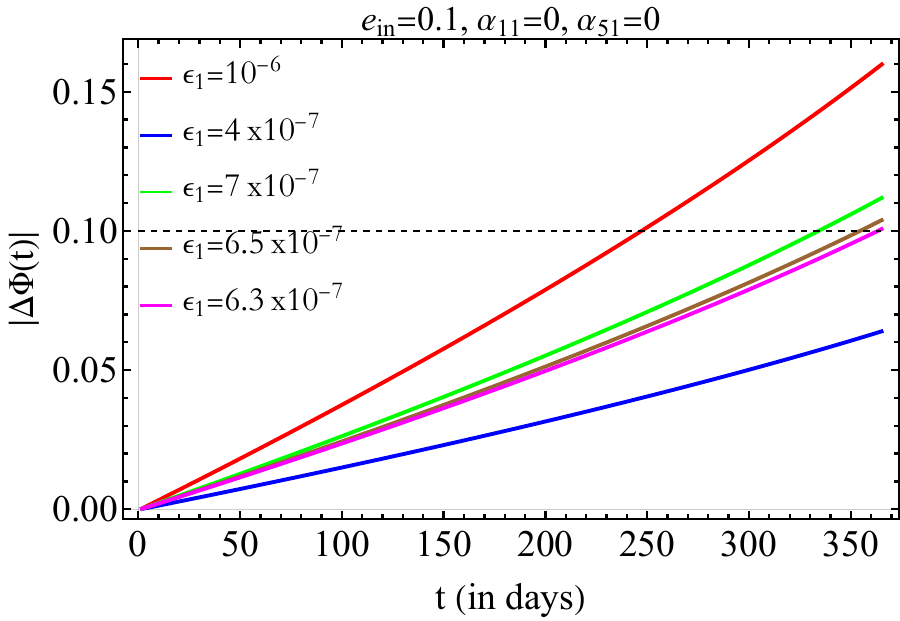}
	 \endminipage
	\caption{Dephasing for a more generic picture of deformed Kerr geometry (VLLC): It provides an order of magnitude estimate of the upper bound on the deviation parameters when $|\Delta\Phi| <0.1$. 
 }\label{constrain_dephase2}
\end{figure}
%%%%%%%%%%%%%%%%%%%%%%%%%%%%%%%%%%%%%%%%%%%%%%%%%%%%%%%

Figs. (\ref{constrain_dephase1}, \ref{constrain_dephase2}) present the order of magnitude analysis in order to detect the deviations with LISA observations. As a thumb rule, we consider the detection threshold with dephasing $\Delta\Phi\gtrsim 0.1$ rad for SNR 30. In both plots, the dotted curve represent the cutoff that enables us to provide an order of magnitude for the detectability of deviations in LISA. For the case of Johannsen in Fig (\ref{constrain_dephase1}), we find that with mass-ratio $q=10^{-5}$ and initial eccentricity $e_{in}=0.1$, the dephasing remains less than 0.1 for deviations $\alpha_{13}\leq 5.5\times 10^{-6}$, $\alpha_{52}\leq 10^{-5}$ and $\epsilon_{3}\leq 10^{-5}$. Therefore, this provides a lower bound on the deviations with the given detection threshold, i.e., if we take these values and below, the detector will not be able to detect the corresponding dephasing. Furthermore, in the case of VLLC, the Fig. (\ref{constrain_dephase2}) again shows the order of magnitude analysis in the similar fashion as presented for the Johannsen spacetime. We notice that the detectors cannot detect the GW dephasing if $\alpha_{11}< 3.5\times 10^{-7}$, $\alpha_{51}<7\times 10^{-7}$ and $\epsilon_{1}<6.3\times 10^{-7}$. Moreover, it is apparent from the plots that the larger values of deviations will lead to larger GW dephasing, making the detectability of deviations more promising with space-based detectors. For a comprehensive details readers are suggested to look at \cite{Kumar:2024utz, AbhishekChowdhuri:2023gvu}.

In order to obtain a more stringent and accurate constraint on deviations we compute mismatch. In principle, one should examine such a case up to 2PN as we have presented the emergence of leading-order deviations (at 2PN) in fluxes and orbital evolution in Johannsen case. However, this require technical advancement in the procedure itself which is also one of focuses of future study in this line. Moreover, we estimate the mismatch for VLLC case where ($\alpha_{11}, \alpha_{51}, \epsilon_{1}$) contribute at the 0PN itself.

\subsection{Mismatch}
In this section, we extend our analysis to compute the mismatch of the GW waveforms, which is supposed to constrain the deviations more stringently. Note that we are only considering the leading-order PN analysis. Hence, we examine the leading-order corrections (deviations) in the detectability. We find that, for the case of Johannsen spacetime (mentioned in section (\ref{JHN})), the leading-order deviations emerge at 2PN. However, in the case of VLLC spacetime (mentioned in section (\ref{sec2})), the leading-order deviations contribute to the 0PN itself. Now, in principle, to perform the waveform/mismatch analysis, one should take into account all effects up to 2PN, including the GR corrections. However, we restate again, this requires several advancements if one is given a black hole background. Therefore, we only focus on the VLLC spacetime for the mismatch computation as it impacts the analysis at the 0PN directly. It is worth adding that we use Eq. (\ref{dpdtn}) for replacing the evolution of ($p(t), e(t)$) in the mismatch computation.

Briefly, the metric perturbation $h_{ij}^{TT}$ represents the gravitational field in the transverse traceless (TT) gauge; the corresponding expression is $h_{ij}^{TT} = \frac{2}{R}\Ddot{I}_{ij}^{TT}$  \cite{Gopakumar:2001dy, Yunes:2009yz, Babichev:2024hjf}.  ($R, I_{ij}$) is the luminosity distance between the detector and source and quadrupole mass moment, respectively. We compute further the polarization modes ($h_{+}, h_{\times}$) in terms of two unit vectors ($p, q$) in the transverse subspace of the propagation direction, with the use of $h_{ij}^{TT}$. Thus the polarization modes take the following form \cite{Moore:2016qxz, Gopakumar:2001dy, Yunes:2009yz},
\begin{align}
h_{+} = \frac{1}{2}(p_{i}p_{j}-q_{i}q_{j})h^{TT}_{ij} \hspace{3mm} ; \hspace{3mm} h_{\times} = \frac{1}{2}(p_{i}q_{j}-p_{j}q_{i})h^{TT}_{ij}\,.
\end{align}
We use the time domain EMRI waveforms to estimate the mismatch ($\mathcal{M}$) to provide an order of magnitude analysis of deviation parameters for observing them with LISA observations. We proceed with defining the overlap between two GW signals ($h_{1}(t), h_{2}(t)$) \cite{PhysRevD.78.124020, Babak:2006uv, Rahman:2022fay}
\begin{align}
O(h_{1}, h_{2}) = \frac{(h_{1}\vert h_{2})}{\sqrt{(h_{1}\vert h_{1})(h_{2}\vert h_{2})}}\,.
\end{align}
Their inner product is given by $(h_{1}\vert h_{2}) = 4Re \int_{0}^{\infty}\frac{\Tilde{h}_{1}(f)\Tilde{h}^{*}_{2}(f)}{S_{n}f}df$.
where $\Tilde{h}(f)$ is the Fourier transform of time domain waveform $h(t)$. The $S_{n}(f)$ is the power spectral density whose explicit expression can be found in \cite{PhysRevD.69.082005, PhysRevD.78.124020, Huerta:2011kt}. Given the overlap, we calculate the Mismatch as 
\begin{align}
\mathcal{M} = 1-O(h_{1}, h_{2}).
\end{align}
If two signals are identical, then $O=1$, as a result, the mismatch will be zero. The $\mathcal{M}$ value lies in the range zero to one.
\begin{figure}[h!]
\centering
\includegraphics[width=2.3in, height=1.6in]{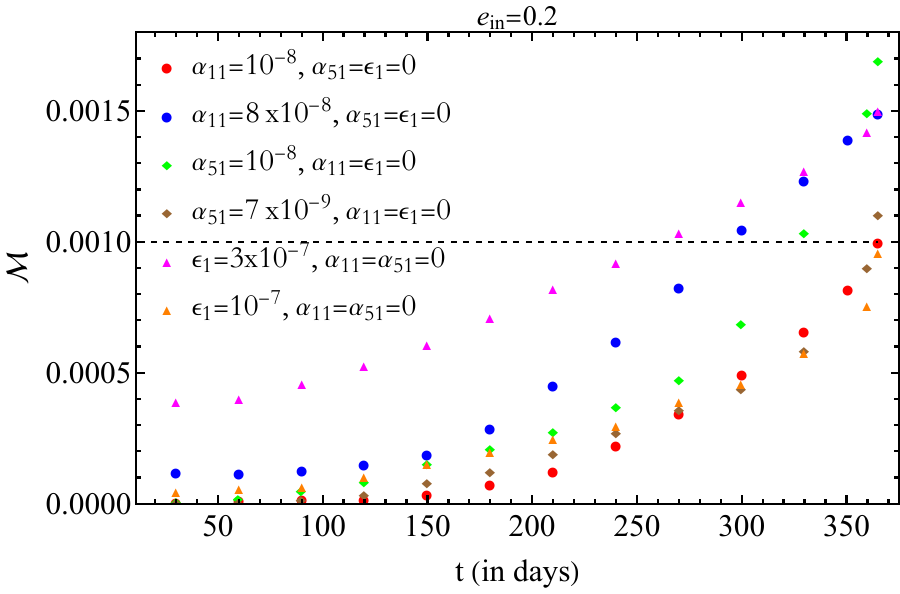}
\includegraphics[width=2.3in, height=1.5in]{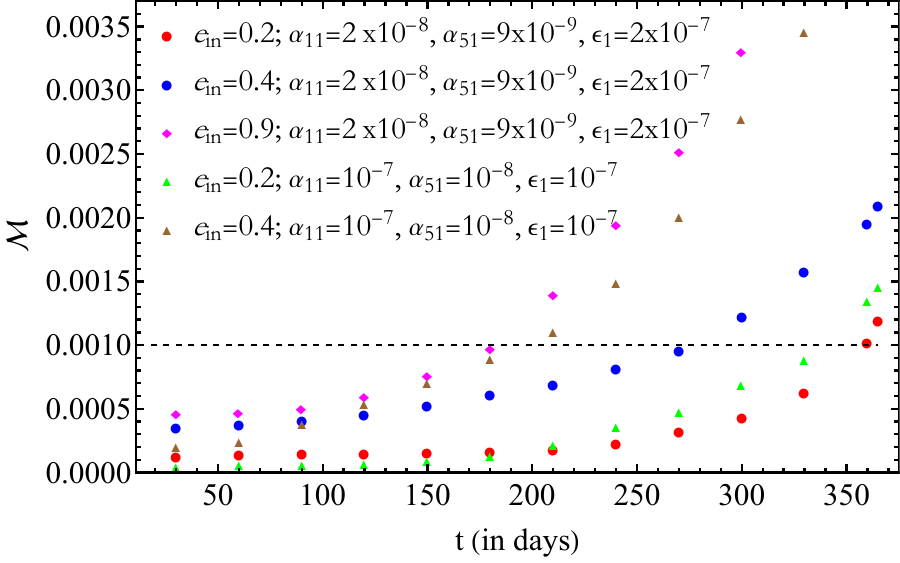}
\caption{VLLC case: mismatch with a primary ($10^{6}M_{\odot}$) and a secondary object ($10M_{\odot}$). It presents an order of magnitude of deviations that enable the estimated mismatch to cross the detection threshold with an observation period of 1 year.} \label{mismatch}
\end{figure}

In Fig.~(\ref{mismatch}), we follow that whenever the mismatch is less than the detection threshold, the waveforms ($h_{1}, h_{2}$) are indistinguishable in GW observations, i.e., $\mathcal{M}(h_{1}, h_{2})\leq \mathcal{M}_{\textup{threshold}} \approx 1/2\rho^{2}$. The $\mathcal{M}_{\textup{threshold}}$ denotes the detection threshold and $\rho$ is the signal-to-noise ratio (SNR). We consider $\mathcal{M}_{\textup{threshold}}\approx0.001$ \cite{PhysRevD.78.124020, Rahman:2022fay, Huerta:2011kt}. With this as a reference, we constrain the deviations in order to detect them in LISA observations. Again, we consider the observation period to be one year. 

The left panel of Fig.~(\ref{mismatch}) depicts the mismatch with an initial eccentricity and setting one deviation parameter non-zero at a time. On the other hand, the right panel of Fig.~(\ref{mismatch}), presents the case where all deviations are turned on simultaneously. In particular, with $e_{\textup{in}}=0.2$, the deviation parameters for which the computed mismatch exceeds the detection threshold are $\alpha_{11} \sim 2\times 10^{-8}$, $\alpha_{51} \sim 9\times 10^{-9}$ and $\epsilon_{1} \sim 2\times 10^{-7}$. It is clear from both plots that the order of magnitude of the deviations remains the same if we keep one deviation non-zero at a time or all non-zero simultaneously.
%%%%%%%%%%%%%%%%%%%%%%%%%%%%%%%%%%%%%%%%%%%%%%%%%%%%%%%%%%%%%%%%%%%%%%%%%%%%%%%%%%%%%%%%%%%%%%%%%%%%%%%%%%%%%%

\section{Discussion}\label{sec5}
In this article, we present an analysis of spacetime that encompasses several deformation parameters, which can be mapped to many other theories beyond GR \cite{Yagi:2023eap}. We perform leading-order PN analysis and examine the effects of leading-order deviations in the GW dephasing and mismatch. We divide the investigation into two categories: first is the case of Johannsen spacetime (based on section (\ref{JHN})), where we impose ppN constraints and notice the emergence of leading-order deviations at the 2PN. Second is the case with VLLC (based on section (\ref{sec2})), where we do not implement any ppN constraints and find the leading-order contribution of deviations at the 0PN itself. Note that one can easily restore Johannsen spacetime by setting the deviation function $A_{0}(r)\longrightarrow A_{1}(r)A_{2}(r)$.

We obtain the analytical expressions of average loss of energy, and angular momentum fluxes with the leading-order PN analysis. We find that the Johannsen spacetime, with the ppN constraints imposed and mass-ratio $q=10^{-5}$ and $e_{in}=0.1$, has a leading-order effect of deviations at the 2PN order, where ($\alpha_{13}, \alpha_{52}, \epsilon_{3}$) are the contributing parameters.  With this, we compute the orbital evolution of the inspiral system and further estimate the GW dephasing as presented in Fig. (\ref{constrain_dephase1}); consequently, we obtain a lower bound on the deviations ($\alpha_{13}\leq 5.5\times 10^{-6}, \alpha_{52}\leq 10^{-5}, \epsilon_{3}\leq 10^{-5}$) considering a detection threshold $\Delta\Phi \gtrsim 0.1$ rad. It implies that if one takes deviations below these values, the corresponding dephasing cannot be detected by the LISA detectors.

On the other hand, in the case of spacetime developed by VLLC, we obtain the contribution of deviations at the 0PN directly, upon relaxing the ppN constraints. The dephasing in this scenario gives an order of magnitude of deviations ($\alpha_{11}< 3.5\times 10^{-7}, \alpha_{51}<7\times 10^{-7}, \epsilon_{1}<6.3\times 10^{-7}$), as presented in Fig. (\ref{constrain_dephase2}), with the corresponding dephasing when the LISA cannot detect such deviations. Further, as mentioned earlier, the complete analysis up to 2PN requires certain advancements, which is also what we are currently working on; hence, we estimate the mismatch for the VLLC case and constrain the deviations more stringently. With the detection threshold $\mathcal{M}\approx 0.001$ at the SNR 30, we obtain a more suitable bound on deviations ($\alpha_{11} \sim 2\times 10^{-8}, \alpha_{51} \sim 9\times 10^{-9}, \epsilon_{1} \sim 2\times 10^{-7}$) in order to detect such deviations with LISA observations, shown in Fig. (\ref{mismatch}). Therefore, as per our investigations, one should consider the values of deviations larger than the ones proposed here as a bound for the observational consequences beyond GR.

Next, we aim to provide a complete study with a 2PN analysis where all deviations contribute to different PN orders up to 2PN. We also aim to conduct the Bayesian analysis to investigate the parameter ranges and how we can study the degeneracy breaking. One can further consider a specific theory and include the non-GR corrections within the fluxes and related quantities, which will further give rise to a more precise bound on the deviations, making more credible detectability of non-GR deviations with LISA observations. We wish to explore some of these aspects in future studies. 

\section*{Acknowledgement}
S. K. gratefully acknowledges the support received for conference travel from the CSIR Travel Grant (TG/12707/24-HRD) and the financial assistance provided by the MG17 conference organizers for accommodation during the event. Additionally, S. K.'s research is funded by the National Post-Doctoral Fellowship (N-PDF) from SERB, DST, Government of India (PDF/2023/000369).  

\bibliographystyle{ws-procs961x669}
\bibliography{JN1}

\begin{thebibliography}{10}
\bibitem{Ghosh:2024arw}
R.~Ghosh and K.~Chakravarti, Parameterized non-circular deviation from the kerr
  paradigm and its observational signatures: Extreme mass ratio inspirals and
  lense-thirring effect, 2406.02454.
\end{thebibliography}

\end{document}